\documentclass[%
 reprint,
 superscriptaddress,
 showpacs,
  nofootinbib,
nobibnotes,
 amsmath,amssymb,
 aps,
   prl,
]{revtex4-1}

\usepackage{verbatim}
\usepackage{graphicx}
\usepackage{epsfig}
\usepackage{amsfonts}
\usepackage{amsmath,amssymb}
\usepackage{bm}
\usepackage[colorlinks=true,citecolor=blue,urlcolor=blue,linkcolor=blue,menucolor=blue]{hyperref}
\usepackage{color}
\usepackage{mathrsfs}

\begin{document}

\title{Prediction for a four-neutron resonance}

\author{A. M. Shirokov}
\email{shirokov@nucl-th.sinp.msu.ru }
\affiliation{Skobeltsyn Institute of Nuclear Physics, Moscow State University,
Moscow, 119991, Russia}
\affiliation{Department of Physics and Astronomy,
Iowa State University, Ames, IA 50011-3160, USA}
\affiliation{Pacific National University, 136 Tikhookeanskaya st., Khabarovsk 680035, Russia}
\author{G. Papadimitriou}
\email{papadimitrio1@llnl.gov}
\affiliation{Nuclear and Chemical Science Division, Lawrence Livermore National Laboratory, Livermore, CA 94551, USA}
\author{A. I. Mazur}
\affiliation{Pacific National University, 136 Tikhookeanskaya st., Khabarovsk 680035, Russia}
\author{I. A. Mazur}
\affiliation{Pacific National University, 136 Tikhookeanskaya st., Khabarovsk 680035, Russia}
\author{R. Roth}
\affiliation{Institut f\"ur Kernphysik, Technische Universit\"at Darmstadt, 64289 Darmstadt, Germany}
\author{J. P. Vary}
\email{jvary@iastate.edu}
\affiliation{Department of Physics and Astronomy,
Iowa State University, Ames, IA 50011-3160, USA}


\begin{abstract}

We utilize various {\em ab initio} approaches to search for a low-lying resonance  in the four-neutron ($4n$) system 
using the JISP16 realistic $NN$ interaction. Our most accurate prediction is obtained using a $J$-matrix extension of the No-Core Shell Model
and suggests a $4n$ resonant state at an energy near~$E_r = 0.8$~MeV with a width of approximately $\Gamma = 1.4$~MeV.


\end{abstract}


\maketitle



With  interest  sparked by a recent experiment \cite{4nexperiment} on the possibility of a resonant four neutron ($4n$) structure (see also \cite{Bertulani2016} for a recent communication)
and while awaiting for 
forthcoming experiments on the same system \cite{GSIteam, Kisamori, Paschalis},  
we search for $4n$ (tetraneutron) resonances using the high precision nucleon-nucleon interaction JISP16 \cite{jisp16}.
The experiment has found a candidate $4n$ resonant state with an energy of 0.83 $\pm$ 0.65 (stat) $\pm$ 1.25 (syst) MeV above the $4n$ disintegration threshold and with an upper limit
of 2.6 MeV for the width. The $4n$ system was probed by studying the reaction between the bound $^4$He nucleus and the weakly bound Helium isotope, $^8$He. It has been shown  \cite{8Hegsm} that the four neutrons in $^8$He form a relatively compact geometry. 
Hence the experimental study of the $\rm{^4He}+{^8He}$ collisions 
is a promising avenue for the isolation of the $4n$ subsystem.  
 
The experimental quest for the very exotic $4n$ structure started almost fifteen years ago when the possibility of a bound $4n$ (or tetraneutron) was proposed \cite{marques2002} in $^{14}$Be
breakup reactions ($^{14}$Be $\to$ $^{10}$Be + $4n$). This experimental result however has not been confirmed. Early calculations of the $4n$ system in a small 
basis \cite{Bevelacqua1980414} found it unbound by about 18.5 MeV.  More recent state-of-the-art theoretical calculations have concluded that without altering fundamental characteristics of the nuclear forces \cite{pieper}, the tetraneutron should not be
bound. More theoretical calculations were performed \cite{bertulani,timofeyuk}, all of them agreeing that a bound tetraneutron is not supported by theory. 
 Calculations performed in the complex energy plane to search of multi-neutron resonances within the Complex Scaling Method \cite{Sofianos,lazauskas,Carbonell} give quantitatively similar results and point to the fact that
the $4n$ 
resonance, if it exists, would have a very  large width (${\sim} 15$~MeV), likely prohibitive for experimental detection. 
The tetraneutron could however exist if confined in a strong external field. In Nature, this would be the case of $^8$He, where the nuclear mean-field is strong enough to confine the tetraneutron around the 
tightly bound 
$\alpha$-core. Once the field is  suddenly removed by knocking out $^4$He, it is expected that the tetraneutron will disintegrate very fast due to its 
anticipated large width. 

There is also a work \cite{Grigorenko} where the continuum response of the tetraneutron was studied. The outcome was that
there exists a resonant-like structure at around 4--5 MeV above threshold, however this structure depends on the 
tetraneutron production reaction mechanism represented by the source term in this study,
and the conclusion was that the $4n$ probably cannot be interpreted as a well-defined resonance but most probably as a few-body continuum response in a reaction.
 
Nevertheless, our current knowledge of nuclear interactions  and  many-body methods provide new opportunities to probe exotic states above thresholds. We are further motivated by the conclusion in Ref.~\cite{pieper} that even though the existence of a bound tetraneutron is ruled out, extrapolations of (artificialy) bound state results to the unbound regime, imply that
there may be a $4n$ resonance at about 2~MeV above the four-neutron threshold. 

A complete investigation of the tetraneutron as a resonant state, would consist of performing calculations of the actual experimental reaction $^4$He($^8$He,$^8$Be). However, such a realistic calculation is currently out of reach, though we are witnessing the first steps for such theoretical calculations to become a reality  \cite{deannature,Quaglioni2015}.

We treat the $4n$ system with a realistic non-relativistic Hamiltonian 
which consists of the kinetic energy and the
realistic inter-neutron potential 
defined by the JISP16 interaction \cite{jisp16}. 
We solve for the $4n$ energies by employing basis expansion techniques for the Hamiltonian. 
Specifically, we employ the No-Core Shell Model (NCSM) \cite{ncsmreview} 
and artificially bind the $4n$ system by scaling the interaction to track its lowest state as a function of that scaling. 
We also employ the No-Core Gamow Shell Model (NCGSM) \cite{ncgsm,reviewgsm} which provides resonant parameters directly in the complex energy plane.
Finally, we extend NCSM using the
Single-State Harmonic Oscillator Representation of Scattering Equations (SS-HORSE) formalism~\cite{EChAYa,5HePRC} for calculations of  the
$S$-matrix resonant parameters.


First, to get an estimate of whether JISP16 can provide a $4n$ resonant state, we exploit the technique suggested in Ref.~\cite{pieper} and
perform pure NCSM calculations by constructing an artificially bound $4n$ system by scaling up the $NN$ interaction. Our extrapolations to the 
unbound regime are in quantitative agreement with  Ref.~\cite{pieper} that predicts a resonance at around 2 MeV above threshold but without 
any indication of the width. We tried also a much more elaborate technique of 
Analytic Continuation in the Coupling Constant (ACCC)~\cite{ACCC, ACCC2}. The ACCC requires exact results for the $4n$ energy with scaled interactions
while NCSM provides only variational energy upperbounds;  extrapolations to the infinite basis space appear to lack the precision needed for
a definite prediction of the resonance energy and width.

In order to shed further light on a possible $4n$ resonance, we solve the NCGSM with the JISP16 interaction.
In the NCGSM one employs a basis set that is spanned by the Berggren states \cite{Berggren1968265} which includes bound, resonant and non-resonant states; they correspond to solutions of the single particle (s.\,p.) Schr\"{o}dinger equation obeying outgoing (bound-resonant states) and scattering (non-resonant states) boundary conditions. In this basis the Hamiltonian matrix becomes complex symmetric and its eigenvalues acquire both real and imaginary parts. The real part is identical to the position of the resonant state above the threshold and the imaginary part is related to its width $\Gamma = -2\,{\rm Im}(E)$.

We adopt the basis provided by a Woods--Saxon (WS) potential for a neutron in relative motion with a $3n$ system. We modify the WS parameters in 
a way that it will support a weakly bound $0s_{1/2}$ state and a resonant $0p_{3/2}$ state. For the $s_{1/2}$ and $p_{3/2}$ shells we include 
the $0s_{1/2}$ bound state, the $0p_{3/2}$ resonant state and the associated non-resonant states. We additionally include the $p_{1/2}$ real 
scattering continuum along the real momentum axis. We performed calculations for several WS parameterizations supporting both narrow and broad s.\,p. states.
States 
with an angular momentum of $\ell$ $>$ 2 are taken as HO states. We retain states through the $3g_{9/2}$ shells. 
For our NCGSM calculations, the $\hbar \Omega$
parameter of the HO basis was varied from 4~MeV to 14~MeV.  Due to the use of Berggren states for low angular momentum partial waves, we observe a weak dependence of the results on the $\hbar \Omega$ parameter. 

For the $4n$ calculation we constructed Slater determinants allowing two neutrons to occupy continuum
orbits, called the 2p-2h approximation. 
Taking the dependence on basis space parameters into account, the NCGSM results indicate a broad resonant state in the energy range 
$E_r\sim 2.5$ to 3~MeV above the $4n$ threshold and a width ranging from $\Gamma\sim 2.5$ to 6~MeV. These variations reflect 
the omission of additional p-h excitations.
Nevertheless the real part of the resonance exhibits a robust character at the current level of p-h truncation, i.\,e., it is nearly independent of the WS parameterizations and independent of the frequency of the HO basis.

At the same time, we observe that the resonance energy decreases together with the width 
as the NCGSM basis increases.
Getting the converged resonance pole position in this approach requires the  NCGSM basis spaces beyond our current reach.
%

Finally, following the $J$-matrix formalism in scattering theory~\cite{Yamani} as 
represented in the HORSE method~\cite{Bang},
we extend the finite NCSM Hamiltonian matrix 
in the harmonic oscillator (HO) basis into the continuum
by appending to it the infinite kinetic energy matrix. 

For the kinetic energy extension of the NCSM Hamiltonian, we use the {\em democratic decay} approximation (also known as {\em true four-body scattering} or {\em 4~$\to$~4 scattering} suggested~\cite{JibKr, JibutiEChaYa})
and first applied to the tetraneutron problem~\cite{Jibuti, jib2, Roman} by Jibuti et al. Later it was
exploited in other tetraneutron studies (see, e.\,g., Refs.~\cite{Efros,Nesterov, Sofianos,Grigorenko}). {Democratic decay} implies a description of the continuum using a complete hyperspherical harmonics (HH) basis. 
In practical applications, 
a limited set of HH is selected which is adequate for the systems like the $4n$ which has no bound subsystems.

The general theory of the {democratic decay} within the HORSE formalism was proposed in Ref.~\cite{SmSh}. We use here
the minimal approximation for the four-neutron decay mode, i.\,e., only HH with hyperspherical momentum~$K=K_{\min}=2$ 
are retained in the kinetic energy extension to the NCSM.
This approximation relies on the fact that the decay in the hyperspherical states
with~$K>K_{\min}$ is strongly suppressed by a large hyperspherical centrifugal barrier~$\frac{\mathscr{L}(\mathscr{L}+1)}{\rho^{2}}$
where the effective momentum~$\mathscr{L}=K+3$ and the hyperradius~$\rho^{2}=\sum_{i=1}^{4}(\mathbf{r}_{i}-\mathbf{R})^{2}$,
$\mathbf{R}$ is the tetraneutron center-of-mass coordinate and~$\mathbf{r}_{i}$ are the coordinates of individual neutrons. 
Note, all possible HH are retained in the NCSM basis.
The accuracy
of this approximation was confirmed in studies of democratic decays in cluster models~\cite{Lur11Li, Lur6He, LurAnn, LurSauAr}.

Realistic $NN$ interactions 
require large NCSM basis spaces and extensive computational resources. 
For computational economy, 
we also adopt the SS HORSE  approach~\cite{EChAYa,5HePRC} where
we calculate the $4\to 4$ $S$-matrix~$S(E)$ at one of the positive eigenenergies of the NCSM Hamiltonian,
$E=E_{\lambda}$. 
In this case, the general HORSE
formula for the $S$-matrix simplifies:
expressing~$S(E)$ through the  $4\to 4$ phase shifts $\delta(E)$,
\begin{equation}
S(E)=e^{2i\delta(E)}, 
\label{S-delta}
\end{equation}
we obtain for the phase shifts~\cite{EChAYa,5HePRC}
\begin{equation}
        \label{SSJM_phase}
        \delta(E_\lambda)=-\tan^{-1}\frac{S_{N^{tot}_{\max}+2,\mathscr{L}}(E_\lambda)}{C_{N^{tot}_{\max}+2,\mathscr{L}}(E_\lambda)}.
\end{equation}
Here the maximal total quanta in the NCSM basis 
$N^{tot}_{\max}=N_{\min}+N_{\max}$, $N_{\min}=2$ is the quanta of the lowest
possible oscillator state of the $4n$ system, 
$N_{\max}$ is the maximal excitation quanta in the NCSM basis; 
analytical expressions for the regular~$S_{N\mathscr{L}}(E)$ and 
irregular~$C_{N\mathscr{L}}(E)$  solutions of the free many-body Hamiltonian in the oscillator representation can be found elsewhere~\cite{SmSh}.
Varying~$N_{\max}$ and~$\hbar\Omega$ in the NCSM calculations, we obtain the phase shifts and $S$-matrix over an energy interval. 
Parametrizing the $S$-matrix in this energy interval, we obtain information about its nearby poles and hence resonances in the system.

\begin{figure}
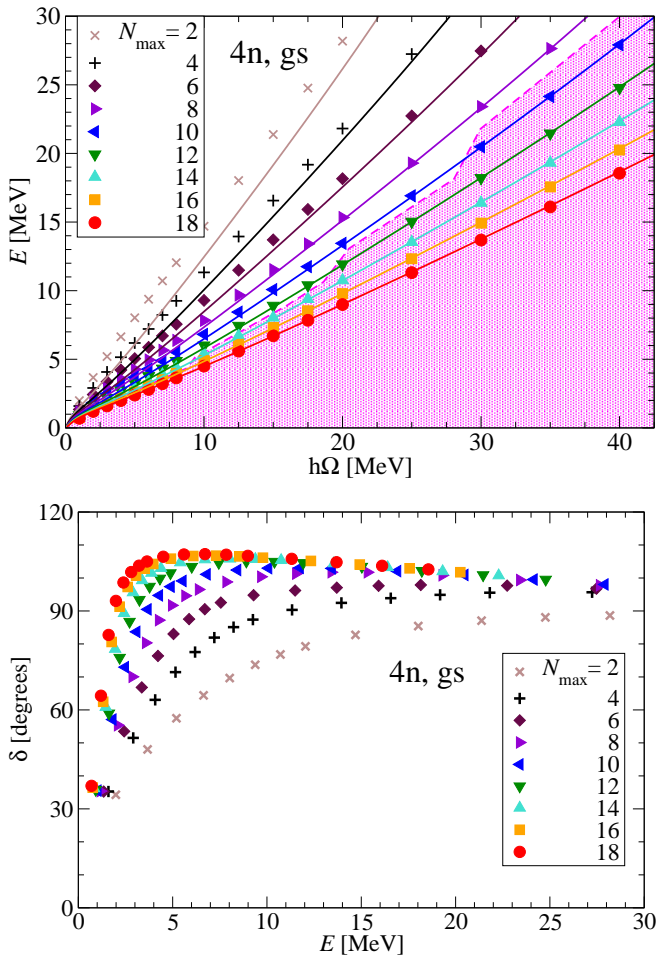

\epsfig{file=4n_Ehw_Select1.eps,width=\columnwidth}\vspace{2ex}
\epsfig{file=4n_deltaE.eps,width= \columnwidth}
\caption{NCSM results for the tetraneutron ground state energy obtained 
with various~$N_{\max}$ (symbols) plotted 
as functions of~$\hbar\Omega$ (upper panel).
The shaded area shows the  NCSM result  selection
for the $S$-matrix parametrization; the solid curves are obtained from the phase shifts 
parametrized with a single 
resonance pole by solving Eq.~\eqref{SSJM_phase} for the eigenenergies at given~$N_{\max}$ and~$\hbar\Omega$ values. 
The $4\to 4$ phase shifts obtained directly from the NCSM results using 
Eq.~\eqref{SSJM_phase}, are shown in the
lower panel.}
\label{NCSM}
\end{figure}

The NCSM calculations were performed with~$N_{\max}=2$, 4, ...\,, 18 
using the
code MFDn \cite{mfdn1,mfdn2} and with $\hbar\Omega$ values,
$\rm 1~MeV\leq\hbar\Omega\leq 40~MeV$. The results for the $0^{+}$ tetraneutron ground state are shown in 
the upper panel of Fig.~\ref{NCSM}. 

The convergence patterns of the NCSM SS-HORSE approach to the $4\to4$ phase shifts using Eq.~\eqref{SSJM_phase} are shown in the lower panel of  Fig.~\ref{NCSM}.
We observe that the phase shifts tend to the same curve when~$N_{\max}$ is
increased. The convergence is first achieved at the higher energies while larger~$N_{\max}$ yield converged
phase shifts at smaller energies. We obtain nearly completely converged phase shifts at all energies with~$N_{\max}=16$ and~18.

\begin{figure}
\epsfig{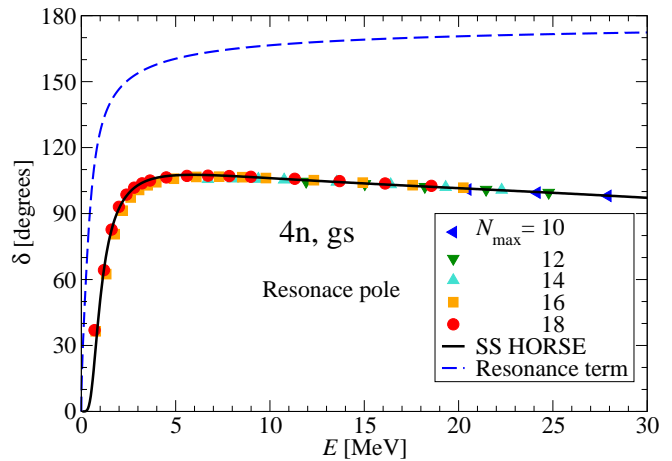}
\caption{The $4\to4$ scattering phase shifts: parametrization with a single 
resonance pole (solid line) and obtained directly from the selected NCSM results using 
Eq.~\eqref{SSJM_phase} (symbols). The dashed line shows the contribution of the resonance term. }
\label{respole}
\end{figure}

\begin{figure}[t!]
\epsfig{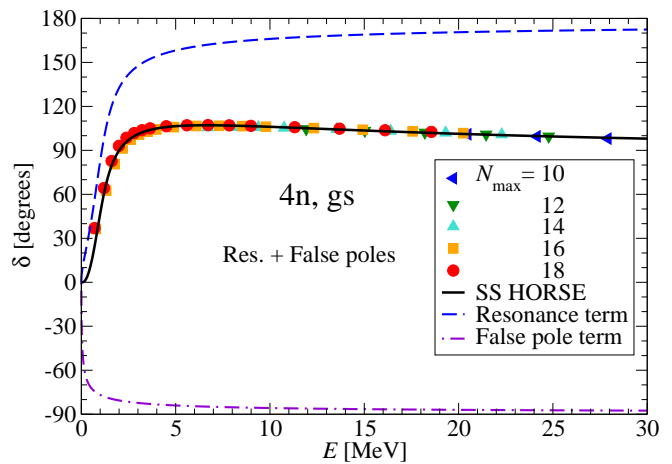}
\caption{Same as Fig.~\ref{respole} but for the  parametrization with   
resonance and false state poles. The dashed-dotted line shows the contribution of the false state pole term.
}
\label{resfalsepole}
\end{figure}

We need only phase shifts close to convergence for the phase shift parametrization.
Our selected NCSM eigenenergies are enclosed by the shaded
area on the top panel of  Fig.~\ref{NCSM} since their 
resulting phase shifts form a single smooth curve (see Figs.~\ref{respole} and~\ref{resfalsepole}).

We will describe now how we utilize the NCSM solutions within the SS-HORSE method in order to obtain resonance positions.
Due to the $S$-matrix symmetry property, $S(k)={1}/{S(-k)}$, and
Eq.~\eqref{S-delta},
the $4\to4$ phase shift~$\delta(E)$ is an odd function of momentum~$k$ and its expansion in
Taylor series of~$\sqrt{E}\sim k$ includes only odd powers of~$\sqrt{E}$:
\begin{gather}
\delta(E)=v_{1}\sqrt{E}+v_{3}\bigl(\sqrt{E}\bigr)^{3}+...+v_{11}\bigl(\sqrt{E}\bigr)^{11}+...
\label{Taylordelta}
\end{gather}
Furthermore, 
the ${4\to4}$ phase shifts at low energies, i.\,e., in the limit~$k\to0$, should behave 
as~$\delta\sim k^{2\mathscr{L}+1}$. Note, in our case, $\mathscr{L}=K_{\min}+3=5$, hence $v_{1}=v_{3}=...=v_{9}=0$
and expansion~\eqref{Taylordelta} starts at the 11$^{th}$ power.

Supposing the existence of a low-energy resonance in the $4n$ system,  we express the $S$-matrix   as $   S(E)=\Theta(E)\,S_r(E),$
where $\Theta(E)$ is a smooth function of energy~$E$ and~$S_r(E)$ is a resonant pole term.  
The respective phase shift is
\begin{equation}
   \delta(E)=\phi(E)+\delta_r(E),
   \label{deltaphip}
\end{equation}
where the pole contribution~$\delta_{r}(E)$ takes the form
\begin{gather} 
\label{deltapE}
\delta_r(E) =- \tan^{-1}\bigl(a\sqrt{E}/(E-b^2)\bigr).
\end{gather}
The resonance energy~$E_r$ and  width~$\Gamma$  are expressed through parameters~$a$ and~$b$ entering Eq.~\eqref{deltapE} as
\begin{equation}
\label{EGab}
E_r 
= b^2-a^2/2,\qquad 
\Gamma=2a\sqrt{b^2-a^2/4}.
\end{equation}
We use the following expression for the
background phase:
\begin{equation}
 \label{Rational3_6}
	\phi(E)=\frac{w_1\sqrt{E}+w_3\bigl(\sqrt{E}\bigr)^3+c\bigl(\sqrt{E}\bigr)^5}{1+w_2E+w_4E^2+w_6E^3+dE^4}.
\end{equation}
The parameters~$w_i,$ $i=1,2,3,4,6$ are uniquely defined through  the parameters~$a$ and~$b$ and guarantee  
the cancellation 
of the terms of powers up to~9 in the 
expansion~\eqref{Taylordelta}.

Our phase shift parametrization is given by Eqs.~\eqref{deltaphip}, \eqref{deltapE} and~\eqref{Rational3_6} with fitting
parameters~$a$, $b$, $c$ and~$d$. For each parameter set, we solve  
Eq.~\eqref{SSJM_phase} to find 
the values of the
energies~$E_{\lambda}^{a,b,c,d}$, and search for the  parameter set~$({a,b,c,d})$ 
minimizing the rms deviation of~$E_{\lambda}^{a,b,c,d}$
from the selected set of NCSM eigenenergies~$E_{\lambda}$.
Following this route, we obtain an excellent description of the selected $E_{\lambda}$
with an rms deviation of~5.8~keV with~$a=0.724$~MeV$^{-\frac12}$, $b^2=0.448$~MeV, $c=0.941$~MeV$^{-\frac52}$, 
and~$d=-9.1\cdot10^{-4}$~MeV$^{-4}$. The resulting predictions for the NCSM eigenenergies are shown by solid lines
in the upper panel of Fig.~\ref{NCSM} where we also describe well  NCSM energies 
with large enough~$N_{\max}$ and/or~$\hbar\Omega$
not included in the minimization fit. We obtain also an excellent description of NCSM-SS-HORSE predicted phase shifts as is shown by solid line in Fig.~\ref{respole}.

However the resonance parameters describing the location of the $S$-matrix pole obtained by this fit, are surprisingly small: the resonance
energy~$E_r=0.186$~MeV and the width~$\Gamma=0.815$~MeV. Note, looking at the phase shift in Fig.~\ref{respole}, we would expect
the resonance at the energy of approximately~0.8~MeV 
corresponding to the maximum of the phase shift derivative 
and
with the width of about~1.5~MeV. The contribution of the pole term~\eqref{deltapE} to the phase shifts is shown by the dashed line in 
Fig.~\ref{respole}. This contribution is seen to differ considerably from the resulting phase shift due to substantial contributions from the 
background phase~\eqref{Rational3_6} which is dominated by the terms needed to fulfill the low-energy theorem~$\delta\sim k^{2\mathscr{L}+1}$
and to cancel low-power terms in the expansion of the resonant phase~$\delta_r(E)$. Such a sizable contribution from
the background in the low-energy region, impels us to search for additional poles or other singularities giving rise to a 
strong energy dependence which would be separate from the background phase.


After we failed to find a reasonable description of the NCSM-SS-HORSE phase shifts with a low-energy virtual state, we found the resolution of the strong background phase problem by assuming that the $S$-matrix has an additional low-energy false pole at a positive imaginary
momentum~\cite{Baz}. We add the false term contribution
\begin{equation} 
\label{PhaseSmatrixFal}
	\delta_f(E)=-\tan^{-1}\sqrt{{E}/{|E_f|}}
\end{equation}
to the phase shift to obtain the equation
\begin{equation}
   \delta(E)=\phi(E)+\delta_r(E)+\delta_f(E)
   \label{deltafalse}
\end{equation}
replacing Eq.~\eqref{deltaphip}.
This parametrization involves an additional fitting parameter~$E_f$. We obtain nearly the same quality 
description of the selected $4n$ ground state energies with the
rms deviation of~6.2~keV with the parameters~$a=0.701$~MeV$^{-\frac12}$, $b^2=1.089$~MeV, $c=-27.0$~MeV$^{-\frac52}$, 
$d=0.281$~MeV$^{-4}$,  and a low-lying false pole at energy~${E_f=-54.9}$~keV. The respective $4n$ resonance at~$E_r=0.844$~MeV 
and width~$\Gamma=1.378$~MeV appears 
consistent with what is expected from directly inspecting the $4n$ phase shifts.
The parametrized phase shifts are shown by solid line
in Fig.~\ref{resfalsepole} together with separate contributions from the resonant 
and false pole 
terms.
We note that corrections introduced by this new parametrization 
to the solid lines in 
Figs.~\ref{NCSM} and~\ref{respole} are nearly unseen in the scales of these
figures.



{\it Conclusions.}
Our results with the realistic JISP16 interaction and the SS-HORSE technique show there is a resonant structure near 0.8 MeV above threshold with a width $\Gamma$ of about 1.4 MeV.
This is the first theoretical calculation that predicts such a low energy $4n$ resonance, without altering any of the properties of the realistic $NN$ interaction.
Our result is compatible with the recent experiment which found a resonant structure at an energy of 0.86 $\pm$ 0.65 (stat) $\pm$ 1.25 (syst) MeV and that set an upper limit for the width at $\Gamma$ = 2.6 MeV. Our complex energy calculations also suggest a broad low-lying $4n$ resonance that agrees marginally with experiment due to the large error bars for both the current application of the NCGSM and the experiment.  

We acknowledge valuable discussions with  Pieter Maris, 
Thomas Aumann, Stefanos Paschalis, Jaume Carbonell and Rimantas Lazauskas. We would also like to thank Nicolas Michel
for sharing the NCGSM code with us.
JPV and AMS thank the Institute for Nuclear Theory at the University of Washington for its hospitality during the completion 
of this work and  Department of Energy for the support of their participation in the INT-16-1 Program.
This work was supported by the US DOE under grants No. DESC0008485 (SciDAC/NUCLEI) and DE-FG02-87ER40371.
This work was also supported by the U.S. Department of Energy, Office of Science, Office of Nuclear Physics, under Work Proposal No. SCW0498 and Award Number DE-FG02-96ER40985.
This work was supported partially through GAUSTEQ
(Germany and U.S. Nuclear Theory Exchange Program for QCD Studies
of Hadrons and Nuclei) under contract number DE-SC0006758.
The development  and application of the SS-HORSE approach was supported by the Russian Science Foundation under
project No.~16-12-10048.
Computational resources
were provided by NERSC, which is supported by the U.S. Department of Energy under Contract No. DE-AC02-05CH11231 and by
Lawrence Livermore National Laboratory (LLNL) institutional Computing Grand Challenge program under Contract No. DE-AC52- 07NA27344. 

%

\begin{thebibliography}{100}


\bibitem{4nexperiment} K. Kisamori {\it et al}, Phys. Rev. Lett. {\bf 116}, 052501 (2016).

\bibitem{Bertulani2016} C. A. Bertulani and V. Zelevinsky,     Nature  {\bf 532}, 448 (2016).


\bibitem{GSIteam} S. Shimoura {\em et al., RIKEN-RIBF proposal on ``Tetra-neutron
resonance produced by exothermic double-charge exchange
reaction,''} NP1512-SHARAQ10.

\bibitem{Kisamori} K. Kisamori  {\em et al., RIKEN-RIBF proposal on ``Many-neutron
systems: search for superheavy $^{7\!}$H and its tetraneutron decay,''}
NP-1512-SAMURAI34.

\bibitem{Paschalis} S. Paschalis, S. Shimoura  {\em et al., RIBF Experimental Proposal},
NP1406-SAMURAI19.

\bibitem{jisp16} A. M. Shirokov, J. P. Vary, A. I. Mazur, and T. A. Weber, Phys. Lett. B {\bf 644}, 33 (2007).

\bibitem{8Hegsm} G. Papadimitriou, A. T. Kruppa, N. Michel, W.~Nazarewicz, M. P\l{}oszajczak, and J. Rotureau, Phys. Rev. C(R) {\bf 84}, 051304 (2011).

\bibitem{marques2002} F. M. Marqu\'es {\it et al} Phys. Rev. C {\bf 65}, 044006 (2002).

\bibitem{Bevelacqua1980414} J. J. Bevelacqua, Nuc. Phys. A {\bf 341}, 414 (1980).

\bibitem{pieper} S. C. Pieper, Phys. Rev. Lett. {\bf 90}, 252501 (2003).

\bibitem{bertulani} C. A. Bertulani and V. Zelevinsky, J. Phys. G {\bf 29} 2431 (2003).

\bibitem{timofeyuk} N. K. Timofeyuk, J. Phys. G {\bf 29}, L9 (2003).

\bibitem{Sofianos} S.~A.~Sofianos, S.~A.~Rakityansky, and G.~P.~Vermaak, J. Phys.
G {\bf 23}, 1619 (1997).

\bibitem{lazauskas} R. Lazauskas and J. Carbonell, Phys. Rev. C {\bf 72}, 034003 (2005).

\bibitem{Carbonell} E. Hiyama, R. Lazauskas, J. Carbonell, and M.~Kamimura, Phys. Rev. C {\bf 93}, 044004 (2016).


\bibitem{Grigorenko} L.~V.~Grigorenko, N.~K.~Timofeyuk,  and M.~V.~Zhukov, Europ. Phys. J.  A {\bf 19}, 187 (2004).

\bibitem{deannature} S. Elhatisari, {\it et al} Nature {\bf 528}, 111 (2015).

\bibitem{Quaglioni2015} S. Quaglioni, Nature {\bf 528}, 42 (2015).

\bibitem{ncsmreview} B. R. Barrett, P.  Navr\'atil, and J. P. Vary, Prog. Part. Nucl. Phys. {\bf 69}, 131 (2013).



\bibitem{ncgsm} G. Papadimitriou, J. Rotureau, N. Michel, M.~P\l{}oszajczak, and B. R. Barrett, Phys. Rev. C {\bf 88}, 044318 (2011).

\bibitem{reviewgsm} N. Michel, W. Nazarewicz, M. P{\l}oszajczak, and T.~Vertse, J. Phys. G {\bf 36}, 013101 (2009).


\bibitem{EChAYa} I. A. Mazur, A. M. Shirokov, A. I. Mazur, and J. P.  Vary, 
arXiv:1512.03983 [nucl-th] (2015).

\bibitem{5HePRC} A. M. Shirokov, A. I. Mazur,  I. A. Mazur, and J. P.  Vary, {\em in preparation}.


\bibitem{ACCC} V. M. Krasnopolsky and V. I. Kukulin, Phys. Lett. A {\bf 69}, 251 (1978).

\bibitem{ACCC2}V. I. Kukulin, V. M. Krasnopolsky, and J. Hor\'acek, {\em Theory of Resonances. Principles and Applications.} (Kluwer, Dordrecht, 1989).

\bibitem{Berggren1968265} T. Berggren, Nucl. Phys. A {\bf 109}, 265 (1968).


\bibitem{Yamani} H.~A.~Yamani and L.~Fishman,  J. Math. Phys.,
{\bf 16}, 410 (1975).

\bibitem{Bang}
J.~M.~Bang, A.~I.~Mazur, A.~M.~Shirokov, Yu.~F.~Smirnov, and
S.~A.~Zaytsev, Ann. Phys. (NY), {\bf 280}, 299 (2000).

\bibitem{JibKr} R. I. Jibuti and N. B. Krupennikova, {\em The Method of Hyperspherical Functions in the Quantum Mechanics of
   Few Bodies} [in Russian] (Metsniereba, Tbilisi, 1984).

\bibitem{JibutiEChaYa} R. I. Jibuti, Fiz. Elem. Chast. At. Yadra {\bf 14}, 741 (1983).


\bibitem{Jibuti} R.~I.~Jibuti, R.~Ya.~Kezerashvili, and K.~I.~Sigua, Yad. Fiz. {\bf 32},
1536 (1980).

\bibitem{jib2} R.~I.~Jibuti, R.~Ya.~Kezerashvili, and K.~I.~Sigua,  Phys. Lett. B
{\bf 102}, 381 (1981).

\bibitem {Roman} R.~Ya.~Kezerashvili, Yad. Fiz. {\bf 44}, 842 (1986) [Sov. J. Nucl. Phys. {\bf 44}, 542 (1986)].

\bibitem{Efros} A.~M.~Badalyan, T.~I.~Belova, N.~B.~Konyuhova, and V.~D.~Efros,
Yad. Fiz. {\bf 41}, 1460 (1985) [Sov. J. Nucl. Phys. {\bf 41}, 926 (1985)].

\bibitem{Nesterov}  I.~F.~Gutich, A.~V.~Nesterov, and I.~P.~Okhrimenko, Yad. Fiz. {\bf 50},
19 (1989).


\bibitem{SmSh}  A. M. Shirokov, Yu. F. Smirnov, and S. A. Zaytsev,
in {\em Modern Problems in Quantum Theory} (Ed. V.~I.~Savrin and
O.~A.~Khrustalev), (Moscow, 1998), p.~184;  S.~A.~Zaytsev, Yu. F. Smirnov, and A. M. Shirokov, 
Teoret. Mat. Fiz. {\bf 117}, 227 (1998)
[Theor. Math. Phys. {\bf 117}, 1291 (1998)].


\bibitem{Lur11Li} Yu.~A.~Lurie, Yu.~F.~Smirnov, and A.~M.~Shirokov,  Izv. Ros. Akad. Nauk, Set. Fiz. {\bf  57}, 193 (1993)
[Bul. Rus. Acad. Sci., Phys. Ser. {\bf 57},  943 (1993)].  

\bibitem{Lur6He} Yu.~A.~Lurie and A.~M.~Shirokov,  Izv. Ros. Akad. Nauk, Set. Fiz. {\bf 61}, 2121 (1997)
[Bul. Rus. Acad. Sci., Phys. Ser. {\bf 61}, 1665 (1997)]. 

 
\bibitem{LurAnn} Yu.~A.~Lurie and  A.~M.~Shirokov,
Ann. Phys. (NY) {\bf 312}, 284 (2004).  


\bibitem{LurSauAr}Yu.~A.~Lurie and  A.~M.~Shirokov,
        In: {\em The  $J$-Matrix Method. Developments and
	Applications.} (Eds. A.~D.~Alhaidari, H.~A.~Yamani, E.~J.~Heller, and 
	M.~S.~Abdelmonem). (Springer, 2008), p. 183.
%
%
%
%
%



\bibitem{mfdn1} P.~Maris, M.~Sosonkina, J.~P.~Vary, E.~G.~Ng, and  C.~Yang,
Procedia Computer Science {\bf 1}, 
97 (May 2010, ICCS 2010).

\bibitem{mfdn2} H.~M. Aktulga, C.~Yang, E.~G.~Ng, P.~Maris, and J.~P.~Vary,
Concurrency Computat. Pract. Exper. {\bf 26}, 2631 (2014).


\bibitem{Baz} A. I. Baz', Ya. B. Zel'dovich, and A. M. Perelomov, {\em Scattering, Reactions and Decay  in Non-Relativistic
Quantum Mechanics.} (Israel Program for Scientific Translation, Jerusalem, 1969).



\end{thebibliography}
%

\end{document}